\documentclass[10pt,conference]{IEEEtran}

\usepackage{cite}
\usepackage{amsmath,amssymb,amsfonts}
\usepackage{algorithmic}
\usepackage{braket}
\usepackage{bbold}
\usepackage{graphicx}
\usepackage{mathtools}
\usepackage{soul}
\usepackage{xcolor}
\def\BibTeX{{\rm B\kern-.05em{\sc i\kern-.025em b}\kern-.08em
    T\kern-.1667em\lower.7ex\hbox{E}\kern-.125emX}}

\newcommand{\alg}{\mathrm{alg}}    
\begin{document}

\makeatletter
\newcommand{\newlineauthors}{%
  \end{@IEEEauthorhalign}\hfill\mbox{}\par
  \mbox{}\hfill\begin{@IEEEauthorhalign}
}
\makeatother

\title{Quantum Resource Estimation for Simulating the SYK Model with Trotterization, qDRIFT, and Asymmetric Qubitization}

\author{\IEEEauthorblockN{Brian Goldsmith}
\IEEEauthorblockA{\textit{Independent Researcher} \\
Buena Park, USA \\
ORCID 0009-0002-5786-4603}
\and
\IEEEauthorblockN{Larissa Kroell}
\IEEEauthorblockA{
\textit{Alice \& Bob}\\
Paris, France \\
ORCID 0000-0001-6965-5652}
\and
\IEEEauthorblockN{Nishna Aerabati}
\IEEEauthorblockA{\textit{University of Illinois Urbana-Champaign} \\
Urbana,  USA \\
nishnaa2@illinois.edu}


}

\maketitle

\begin{abstract}
The Sachdev-Ye-Kitaev (SYK) model has been identified as a promising candidate to run on early fault-tolerant quantum computers due to the relatively modest resources required to probe non-trivial physics (namely holographic duality and AdS/CFT correspondence). As such, it is crucial that the details of how to run such a simulation are well understood. Using PsiQuantum’s Construct platform, we implement and analyze three different approaches to simulate the SYK model: Trotterization, qDRIFT, and asymmetric qubitization with Quantum Signal Processing. We provide an open-source library containing implementations for SYK simulation using all three methods, which we use to obtain quantum resource estimates for qubit and T gate count as functions of the number of Majorana modes and precision. We find that while qDRIFT and Trotterization benefit from a lower qubit count, the large number of T gates required lead to asymmetric qubitization being advantageous in most cases. This reinforces previous theoretical considerations. We intend both the implementations and the estimates to be useful for researchers to continue to study the SYK model and understand how the techniques and resources vary.

\end{abstract}


\section{Introduction}
The dynamics of a physical system give key insights to understanding its properties, but for quantum systems, understanding how the system evolves in time can be a difficult and computationally intensive task. Quantum computers have the potential to overcome these challenges, but many relevant physical systems require quantum computers that are larger than those currently available. As progress continues towards larger quantum computers and fault-tolerant quantum computing, it is important to distinguish small and intermediate scale models as potential first candidates to run on early fault-tolerant quantum computers. By studying these models, researchers can develop and test workflows for running quantum simulations.  This allows for not only resource estimation to verify asymptotic costs, but also thorough testing to validate that the program is runnable on hardware.

The  Sachdev-Ye-Kitaev (SYK) model is an ideal candidate for such studies. The SYK model is an exactly solvable quantum many-body system which also exhibits holographic duality \cite{Polchinski2016}. Specifically, Kitaev proposed the model as a holographic dual to a black hole in the anti-de Sitter (AdS) spacetime \cite{kitaev_video}. 

At finite temperatures, the SYK model exhibits maximal chaos, characterized by a Lyapunov exponent that saturates the upper bound allowed by large-N constraints \cite{kitaev_video}. This maximal chaos is a property that is also exhibited by black holes in Einstein gravity \cite{Maldacena2016}. Consequently, the SYK model is an example of the general AdS/CFT correspondence, which is a conjectured duality between coupled gauge theories and gravitational theories on the AdS spacetime \cite{Maldacena1998}. While these holographic properties are mathematically defined, their physical realization on quantum hardware depends on the ability to simulate the model's high-dimensional interaction space. Furthermore, although the SYK model is exactly solvable in specific limits, important properties, such as the density of states, remain analytically challenging \cite{Asym_Qubitization}, further motivating quantum simulation approaches.

There are multiple ways to simulate the SYK model on a quantum computer, and different approaches lead to vastly different resource requirements. To evaluate these requirements we use the following definitions for the SYK model.  The Hamiltonian for the SYK model we consider is given as follows \cite{kitaev_video}:
\begin{equation}\label{eq:syk_ham_1}
    H_\mathrm{SYK} = \frac{1}{4 \cdot 4!} \sum_{p,q,r,s}^N J_{pqrs}\chi_p\chi_q\chi_r\chi_s,
\end{equation}
where $J_{pqrs}$ are Gaussian coefficients dependent on a given coupling constant $J$. Here, $\chi_p$ denote  Majorana fermions, i.e., $\chi_p$ is Hermitian and pairs satisfy the known anti-commutator relation  $\{ \chi_p, \chi_q \} = 2 \delta_{pq}$. 

Note that the notation used in Eq.~\eqref{eq:syk_ham_1} is ambiguous. If we enforce antisymmetry in the coefficients $J_{pqrs}$, we obtain a quartic Hamiltonian, since terms with repeated indices (e.g. $J_{pprs}$ etc.) must be zero. However, we can also treat the coefficients as independent and identically distributed (i.i.d.) random variables, leaving the commutation relations of the Majorana operators themselves to generate the required symmetries. Doing this results in a Hamiltonian with both quartic and quadratic terms, which exhibits slightly different physics. Anticipating that the antisymmetrized Hamiltonian will primarily be used here for Trotterization-based methods, we write
\begin{equation}
    H_\mathrm{SYK}^\mathrm{trot} = \frac{1}{4 \cdot 4!} \sum_{p< q< r < s}^N J_{pqrs}\chi_p\chi_q\chi_r\chi_s,\label{eq: H SYK trot}
\end{equation}
where $J_{pqrs} \sim \mathcal{N}\big(0,\tfrac{3! \Tilde{J}^2}{N^3}\big)$ are independent Gaussian random variables and $\Tilde{J} = 4! J$ denotes the rescaled coupling constant obtained by taking Majorana operator reordering into account. The Hamiltonian with all i.i.d. random coefficients $J_{pqrs}$ used for the asymmetric qubitization is then
\begin{equation}
    H_\mathrm{SYK}^\mathrm{asym} = \frac{1}{4 \cdot 4!} \sum_{p,q, r,  s = 1}^N J_{pqrs}\chi_p\chi_q\chi_r\chi_s,\label{eq: H SYK qubit}
\end{equation}
with $J_{pqrs} \sim \mathcal{N}\big(0,\tfrac{3! J^2}{N^3}\big)$. It is worth noting that Trotter-based methods are more flexible and can in principle handle either Hamiltonian without modifications, whereas asymmetric qubitization would necessitate some adjustments.

The rescaled coupling constant for $H_\mathrm{SYK}^\mathrm{trot}$ is important to ensure that the coefficients for the four-body terms in both $H_\mathrm{SYK}^\mathrm{trot}$ and $H_\mathrm{SYK}^\mathrm{asym}$ share the same distribution.  This follows from
\begin{equation*}
    \sum_{p,q, r,  s = 1}^N J_{pqrs}\chi_p\chi_q\chi_r\chi_s = \sum_{p< q< r < s}^N \Tilde{J}_{pqrs} \chi_p\chi_q\chi_r\chi_s,
\end{equation*}
where $\Tilde{J}_{pqrs}$ is a linear combination of all $J_{pqrs}$ taking into account all possible indices permutations. Now, since a linear combination of independent normally distributed random variables with zero mean is again normally distributed with zero mean and linearly scaling variance, $\Tilde{J}_{pqrs} \sim \mathcal{N}\big(0,\tfrac{3! (4!J)^2}{N^3}\big)$ as claimed above. However, due to the presence of quadratic terms in $ H_\mathrm{SYK}^\mathrm{asym} $, the two Hamiltonians still differ. For more information on these different instances of the SYK model, see the supplemental material of \cite{SYK_trotter}.

\section{Simulation Techniques}
We implement three different methods to simulate the time evolution for a system evolving according to Eq.~\eqref{eq:syk_ham_1}. The first two are based on Pauli Product Rotations (PPRs), $e^{i\theta P}$, for a Pauli string $P$ \cite[Section 4.7.3]{Nielsen_Chuang_2010}, and act on a shared implementation of the Hamiltonian given in Eq.~\eqref{eq: H SYK trot}, whereas the third approach is based on a block encoding method leading to the implementation of the Hamiltonian given in Eq.~\eqref{eq: H SYK qubit}. We give a brief overview of the methods below.

Generally, for Hamiltonian simulation it is beneficial to expand the Hamiltonian $H$ in terms of simpler unitaries, i.e.
\begin{equation}\label{eq: Hamiltonian decomp}
    H = \sum_{\ell = 1}^L w_\ell H_\ell,
\end{equation}
where $H_\ell$ is often assumed to be given by a Pauli string.
For example, $H_\mathrm{SYK}$ is already given as a sum of products of Majorana modes. However, in order to encode this on a quantum computer, we would like to express this as a Pauli string using the Jordan-Wigner transform \cite{JW1928}.  Depending on the simulation method, we use slightly different conventions. Explicitly, for PPR based methods we use the following conventions
\begin{equation}\label{eq: JW compact}
\begin{aligned}
    \chi_{2\ell} &= \prod_{k=0}^{\ell-1} Z_k X_\ell, \qquad \chi_{2\ell + 1} &=  \prod_{k=0}^{\ell-1} Z_k Y_\ell,
\end{aligned}
\end{equation}
for $\ell = 0, \ldots, N/2-1$, whereas for asymmetric qubitization we use
\begin{equation}\label{eq: JW asym qubit}
    \chi_\ell = \prod_{k=0}^{\ell-1} Z_k X_\ell
\end{equation}
for $\ell = 0, \ldots, N-1$. Observe that in Eq.~\eqref{eq: JW compact}, we need $N/2$ qubits to represent $N$ Majorana modes, whereas for Eq.~\eqref{eq: JW asym qubit} we need as many qubits as Majorana modes.


\subsection{Trotterization} 
Trotterization \cite{SYK_trotter} approximates the unitary evolution $U(t) = e^{-iHt}$ by decomposing the Hamiltonian into a sequence of executable gates. Given a Hamiltonian as defined in Eq.~\eqref{eq: Hamiltonian decomp}, for a total evolution time $t$ divided into $r$ Trotter steps of size $\Delta t = t/r$, the second-order Trotter-Suzuki decomposition for a single time step $\Delta t = t/r$ is given by:
$$S_2(\Delta t) = \prod_{\ell=1}^L e^{-i w_\ell H_\ell \frac{\Delta t}{2}} \prod_{\ell=L}^1 e^{-i w_\ell H_\ell \frac{\Delta t}{2}}.$$
The full time-evolution operator is then obtained by sequentially applying this approximate propagator $r$ times \cite{PhysRevX.11.011020},
\begin{equation}
U(t) = \left[S_2\left(\tfrac{t}{r}\right)\right]^r + \mathcal{O}\left(\frac{t^3}{r^2}\right),
\end{equation}
where the error arises from the non-commutativity of the Hamiltonian terms.
The SYK model's four-body interactions scale combinatorially with the system size $$L = \binom{N}{4}  \approx \mathcal{O}(N^4),$$ where $L$ is the number of terms. For a single $S_2$ Trotter step, each of the $L$ terms must be evolved forward and backward, requiring approximately $2L$ PPRs. Consequently, the total rotation gate count, $R$, scales as $R \approx 2 r  \binom{N}{4}$, resulting in an overall cost of $\mathcal{O}(r N^4)$. This scaling presents a significant computational challenge for intermediate-scale devices. 

To determine the number of steps, $r$, necessary to achieve a target precision $\varepsilon_\alg$, we use the commutator-based error bound (Theorem 6 in \cite{PhysRevX.11.011020}). For the Hamiltonian given in Eq.~\eqref{eq: H SYK trot}, the commutator scaling factor is given by, 
\begin{equation}
\tilde{\alpha}_{\text{comm}} = \sum_{\ell_1, \ell_2, \ell_3 = 1}^L \left\| [H_{\ell_3}, [H_{\ell_2}, H_{\ell_1}]] \right\|,
\end{equation} where $H_\ell$ denotes an individual weighted Hamiltonian term in the decomposition given in Eq.~\eqref{eq: Hamiltonian decomp}. This bound is tighter than one based on the 1-norm, as it accounts for the anti-commutation relations of the Majorana operators $\{ \chi_p, \chi_q \} = 2 \delta_{pq}$ \cite{Maldacena2016}. For a second-order formula ($p=2$), the required number of steps scales as
\begin{equation}
r \approx \sqrt{\frac{\tilde{\alpha}_{\text{comm}} t^3}{\varepsilon_\alg}}.
\end{equation}
The rapid increase in complexity highlights the necessity for the randomized or block encoding techniques, discussed in the subsequent sections.

\subsection{qDRIFT} Instead of performing a sequence of PPR compositions deterministically, relying on the approximation of the time evolution using the Trotter-Suzuki decomposition, the qDRIFT framework \cite{qdrift} randomly samples corresponding PPRs. More specifically, consider a Hamiltonian as in Eq.~\eqref{eq: Hamiltonian decomp} with $w_\ell\geq 0$ for all $\ell$ and let $\lambda = \sum_{\ell=1}^N w_\ell$ denote the 1-norm of the Hamiltonian. Then, if we want to simulate $e^{-iHt}$ for some time $t$, we instead compute $N$ factors $e^{-iH_\ell\frac{t}{N}}$, where $H_\ell$ is sampled from the probability distribution giving weight $w_\ell/\lambda$ to summand $H_\ell$. 

Instead of determining the correct number of Trotter steps, the question becomes how many samples to draw. When qDRIFT was first introduced, it was shown that to achieve an error of $\varepsilon_\alg$ in terms of the diamond norm of the corresponding unitary channels, it suffices to compute $n = \big\lceil \frac{2 (t\lambda)^2}{\varepsilon_\alg}\big\rceil$ samples. Recently \cite{qdriftbettererror}, this has been improved to a linear dependence on $\lambda$ to $n = \big\lceil \frac{4 \lambda t^2}{\varepsilon_\alg}\big\rceil$. Regardless of the power, the key observation here is that the required sample size does not depend explicitly on the number of terms, $L$, in the Hamiltonian. Instead, this dependency only enters implicitly through the norm of the Hamiltonian. For example, since $J_{pqrs}$ are i.i.d. random variables for $H_\mathrm{SYK}^\mathrm{trot}$ as in Eq.~\eqref{eq: H SYK trot}, the expected value of the norm of  $H_\mathrm{SYK}^\mathrm{trot}$ is given by
\begin{equation*}
    \mathbb{E}[\lambda] = \sum_{p<q<r<s}^N \mathbb{E}[\vert J_{pqrs}\vert] = \binom{N}{4} \mathbb{E}[\vert J_{1234}\vert]
\end{equation*}
making the dependency on the number of terms explicit.

\subsection{Asymmetric Qubitization and QSP}  
Asymmetric qubitization\cite{Asym_Qubitization} is a modified version of qubitization\cite{qubitization} which is a method of block encoding the Hamiltonian into a unitary operation to produce a quantum walk operator. Qubitization consists of the three steps, PREPARE, SELECT, and UNPREPARE, followed by a reflection. The produced quantum walk  has eigenvalues of $e^{\pm i \arccos(\frac{h}{\lambda})}$ for eigenvalue $h$ of $H$ and a normalization factor $\lambda$.
Standard qubitization block encodes the Hamiltonian using state preparation causing the amplitudes to be proportional to the square roots of the coefficients, $w_\ell$ from Eq.~\eqref{eq: Hamiltonian decomp}, resulting in
\begin{equation}
    \bra{G}U\ket{G} = \frac{H}{\lambda_1}.
\end{equation}
However, asymmetric qubitization uses two state preparation oracles with the unitary V to produce
\begin{equation}
    \bra{B}V\ket{A} = \frac{H}{\lambda_2}.
\end{equation}
It is important to note that while for conventional qubitization $\lambda_1$ is a normalization factor corresponding to the 1-norm of the Hamiltonian, for asymmetric qubitization $\lambda_2$ is a normalization factor that appropriately scales the results of PREPARE to match the Hamiltonian, including addressing the variance scaling from the Gaussian-like distribution in Oracle A.

For asymmetric qubitization of the SYK model, PREPARE combines two state preparation oracles, Oracle and Oracle B. Oracle A consists of a random orthogonal circuit to produce Gaussian-like distributed amplitudes, and Oracle B applies Hadamard gates, resulting in
\begin{equation}
    \frac{H}{\lambda_2} = \bra{0}B^\dagger U A\ket{0} = \sum_{l=0}^{L-1}\alpha_\ell \beta^*_\ell H_\ell.
\end{equation}
The two oracles are relatively inexpensive and are able to produce the coefficients proportional to the Gaussian distribution much more efficiently than the standard state preparation required by symmetric qubitization. The efficient state preparation makes up for the additional requirements needed by asymmetric qubitization, such as the auxiliary qubits needed for controlling the two PREPARE oracles and the SELECT unary iteration, as well as the additional complexity from using $\lambda_2$ which \cite{Asym_Qubitization} shows is only 1.25. 

As previously mentioned, the unitary produced from PREPARE, SELECT, and UNPREPARE is combined with a reflection to produce a quantum walk, $W$. By interspersing rotations with controlled $W$ and $W^\dagger$, Quantum Signal Processing (QSP) is used to simulate the time-evolution of the Hamiltonian.

\section{Methods}

\subsection{Implementation}
\begin{figure}
    \centering
    \includegraphics[width=\linewidth]{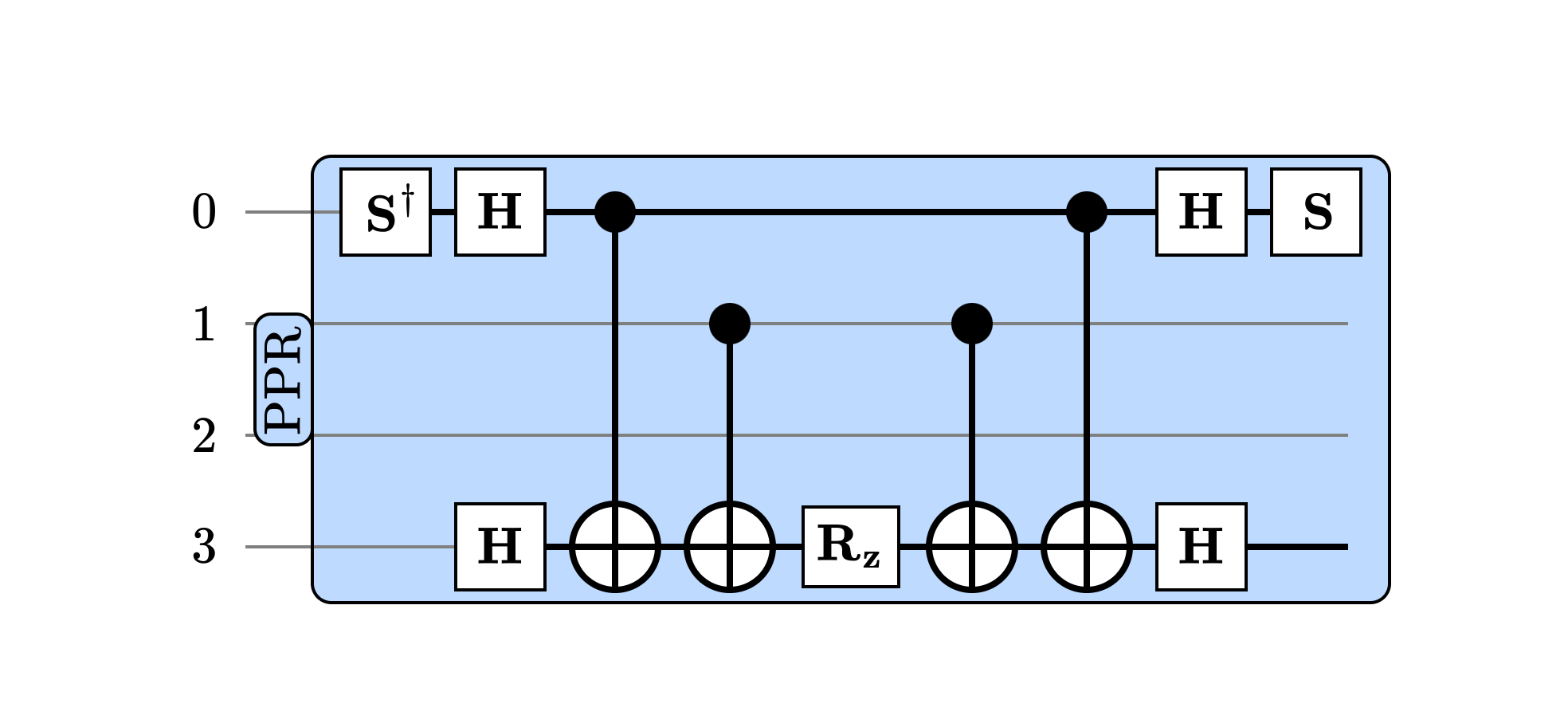}
    \caption{Circuit diagram for Pauli Product Rotation $e^{i\theta Y_0 Z_1 X_3}$ obtained using Construct's Circuit Designer. The $R_z$ rotation in the diagram performs a rotation with angle $2\theta$.}
    \label{fig:PPR_circ}
\end{figure}

We implement the algorithms in Workbench using PsiQuantum's Construct platform. The code, Jupyter Notebooks, circuit diagrams, and resource analysis files are available as an open-source library on GitHub \cite{GitHubRepo}. 

Since Trotterization and qDRIFT are both based on PPRs, they share several algorithmic building blocks. Specifically, both use a Qubrick implementation of a PPR. Figure \ref{fig:PPR_circ} shows a sample circuit for the implemented PPR corresponding to $\chi_0\chi_4\chi_5\chi_6$, which equals $Y_0Z_1X_3$ after using the compact Jordan-Wigner transform.  As both simulation methods apply to more general Hamiltonians, we implement and test them in a more general setting.

For the simulation of the SYK model, we combine the general Trotter and qDRIFT implementations with an expression of $H_\mathrm{SYK}^\mathrm{trot}$ as a sum of Pauli strings using Workbench's PauliSum object. This is achieved by applying the Jordan-Wigner transform to pairs of Majorana modes and constructing the four-body modes through their products, while explicitly keeping track of any incurred phases. This method is drastically simplified by only considering indices of increasing order as stated in Eq.~\eqref{eq: H SYK trot}.

The implementation of the asymmetric qubitization technique from \cite{Asym_Qubitization} differs from the other two techniques by using all combinations of the indices and using a different Jordan-Wigner representation. The focus of \cite{Asym_Qubitization} is the asymmetric qubitization technique, so both the use of all indices and the transformation were chosen for simplicity and still give an upper bound for the complexity of simulating the SYK Hamiltonian stated in Eq.~\eqref{eq: H SYK trot}.
For Oracle A of PREPARE, the Gaussian distribution is produced using layers of CNOTs and random orthogonal rotations, using $R_y$ gates to ensure real coefficients. The number of layers is calculated by doubling the size of the index register, $2\log_2(4^N)$. This is combined with Oracle B which only consists of applying Hadamards to the index register. For SELECT, unary iteration is used to perform indexed X operations and range Z operations as a streaming Jordan-Wigner transformation. After a NOT on the branch qubit and UNPREPARE, the reflection completes the quantum walk.

After generating the quantum walk, the python library \textit{pyqsp}\cite{cdghs_finding_qsp_angles_20,dmwl_efficient_phases_21,mrtc_unification_21} is used to obtain the rotations needed to implement a polynomial approximation of the cosine and sine components of the Hamiltonian evolution. With the quantum walk operator and the rotation angles, QSP is then performed by alternating rotations with $W$ or $W^\dagger$.  The technique from \cite{mrtc_unification_21} reduces the number of Walk operations and therefore T gates. However, it is necessary to introduce two auxiliary qubits for QSP, one as the signal qubit and the second to be the selection qubit that controls applying either the sine-based rotation or the cosine-based rotation. The selection qubit is also used for postselection since it must be measured in the $\ket{0}$ state for the signal processing to be successful.

\subsection{Verification and Validation}
When implementing the techniques, it is critical to ensure  implementation correctness and scientific fidelity. Therefore, comprehensive testing is written to confirm the techniques are implemented properly according to literature. Unfortunately,
validation testing, like level-spacing statistics \cite{rratio, syk_rratio}, for the techniques is inconclusive since the test systems are too small to reveal the chaotic signatures of the SYK model.

The testing for Trotter and qDRIFT again share some similarities. First, the shared PPR implementation is cross-validated using Workbench's implementation. To do so, we compare the corresponding unitaries using Workbench's ``unitary filter'' for smaller systems and for larger systems rely on statevector comparisons. Both Hamiltonian simulation methods are then validated using unit tests regarding error convergence and comparisons with exact results for simple Hamiltonians such as the Heisenberg or Transverse field Ising model. Additionally, we assert that our Trotter implementation matches the implementation in Workbench. The implementation of the Hamiltonian is tested using a range of edge cases as well as validation of swap symmetry. After combining Trotter and qDRIFT with the implementation of the SYK model, we assert the correct behavior on a four qubit system, i.e., eight Majorana modes, of both methods by computing the fidelity $\tfrac{2}{N} \mathrm{Tr} (U_1^\dagger U_2)$, where $U_1$, $U_2$ denote the corresponding unitaries of the quantum algorithm.

For asymmetric qubitization, tests verify each component separately, as well as the complete QSP process. The tests for the random quantum circuit produced in Oracle A include verifying that the mean is zero, the variance is $1/2^M$, where $M$ is the length of the index register, and the maximum amplitude does not dominate the distribution. Since Oracle B consists of only Hadamards, a simple test confirms the amplitudes are set to $1/\sqrt{2^M}$. For SELECT, which applies the Jordan-Wigner transformation, randomly generated indices $(p,q,r,s)$ are passed to SELECT and are also used to manually perform the transform by explicitly calling the appropriate gates. The two resulting states are then compared to verify the SELECT transformation behavior. The full block-encoded unitary is also tested for $N=4$ and $N=8$. The coefficients are used along with 2x2 matrices of $X$ and $Z$ to produce a matrix realization of the unitary operation. Then computational-basis probing is used to produce a similar matrix representation from the circuit for comparison against the matrix realization. For the full QSP process, a test verifies the state fidelity using the unitary matrix realization. 

\subsection{Resource Estimation}
The Workbench framework includes resource estimation using QPU filters. The ``witness'' filter is the primary filter for keeping track of resources required to execute a routine, including gate counts like Gidney elbows\cite{gidney_halving_2018}, Toffolis, rotations, T gates, and measurements. The ``witness'' filter also determines other requirements, such as highwater qubit count, active volume, and more. For our experiments, we focus on T gates and the highwater qubit count. Workbench conveniently includes additional filters that help decompose more complex gates. We use the ``clean-ladder-filter'', which decomposes multi-control gates into Toffoli and rotation gates, the ``single-control-filter'', which decomposes single control gates to CNOT and one qubit gates, and finally ``toffoli-filter'', which decomposes Toffoli gates into Clifford + T gates. With these filters in place, our routines are compiled down to T gates, rotations, and Clifford gates. While Workbench does include an ``rs-synth-filter'' that approximates rotation synthesis, we use the mean linear fit of the cost for Clifford+T using mixed fallback from \cite{rotation_synthesis_eq} to convert the number of rotations to T gates for a final count. 

To compute the necessary error for the rotational synthesis, we assume an additive total error of 
\begin{equation*}
\varepsilon_\mathrm{total} = \varepsilon_\mathrm{rot} + \varepsilon_\mathrm{alg},
\end{equation*}
where $\varepsilon_\mathrm{alg}$ denotes the error from the different simulation methods described in Section II and $\varepsilon_\mathrm{rot}$ denotes the total error from performing rotations on a fault-tolerant quantum computer. For simplicity, we assume $\varepsilon_\mathrm{alg} = \varepsilon_\mathrm{rot}$. For a given $\varepsilon_\mathrm{rot}$, we can then compute the necessary tolerance $\varepsilon_\mathrm{syn}$ for the rotation synthesis by again assuming a linear accumulation of errors
\begin{equation*}
    \varepsilon_\mathrm{rot} \approx N_\mathrm{rot} \varepsilon_\mathrm{syn},
\end{equation*}
where $N_\mathrm{rot}$ denotes the number of rotations in a given circuit. In particular, for fixed $\varepsilon_\mathrm{rot}$ the three simulation techniques require a different $\varepsilon_\mathrm{syn}$ and thus lead to a distinct T count per rotation.

For qDRIFT and Trotterization, most of the resource estimates are based on theoretical worst case estimates, determined by the upper bounds on the error estimates described in Section II-A and II-B. This simpler method is used since the only non-Clifford gate necessary for PPR based methods is given by one rotation per applied PPR. Using analytical bounds circumvents the high computational cost of direct simulation. For small system sizes the rotation counts were verified using Workbench's QPU simulator together with the ``witness'' filter mentioned above.


\section{Results and Discussion}
In this section, we discuss and compare algorithmic scaling in terms of T gates and qubit count for all three methods. To ensure the same distribution of four-body Majorana terms, we use the scaled coupling constant $J=24$ for qDRIFT and Trotter, whereas for asymmetric qubitization, the equivalent coupling constant of $J=1$ is used. All simulations are executed for $t=1$.

\begin{figure}
    \centering
    \includegraphics[width=1\linewidth]{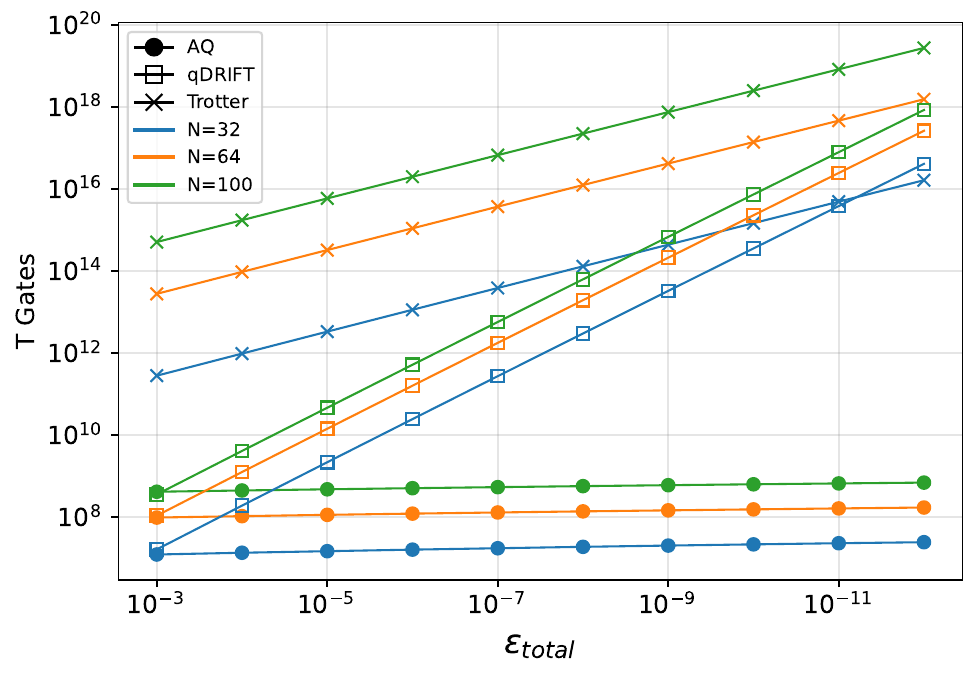}
    \caption{Number of $T$ gates as a function of the total error $\varepsilon_\mathrm{total}$ for different methods at different Majorana counts. While qDRIFT requires fewer T gates for large $\varepsilon_\mathrm{total}$, asymmetric qubitization is better overall with near constant scaling with $\varepsilon_\mathrm{total}$. When calculating the T gates, $\varepsilon_\mathrm{rot}$ is assumed to be equal to $\varepsilon_\mathrm{alg}$, as described in Section III-B. The coupling constants are set to equivalent values ($Jt = 24$ for Trotter and qDRIFT and $Jt=1$ for AQ). }
    \label{fig:Escaling}
\end{figure}
\begin{figure}
    \centering
    \includegraphics[width=1\linewidth]{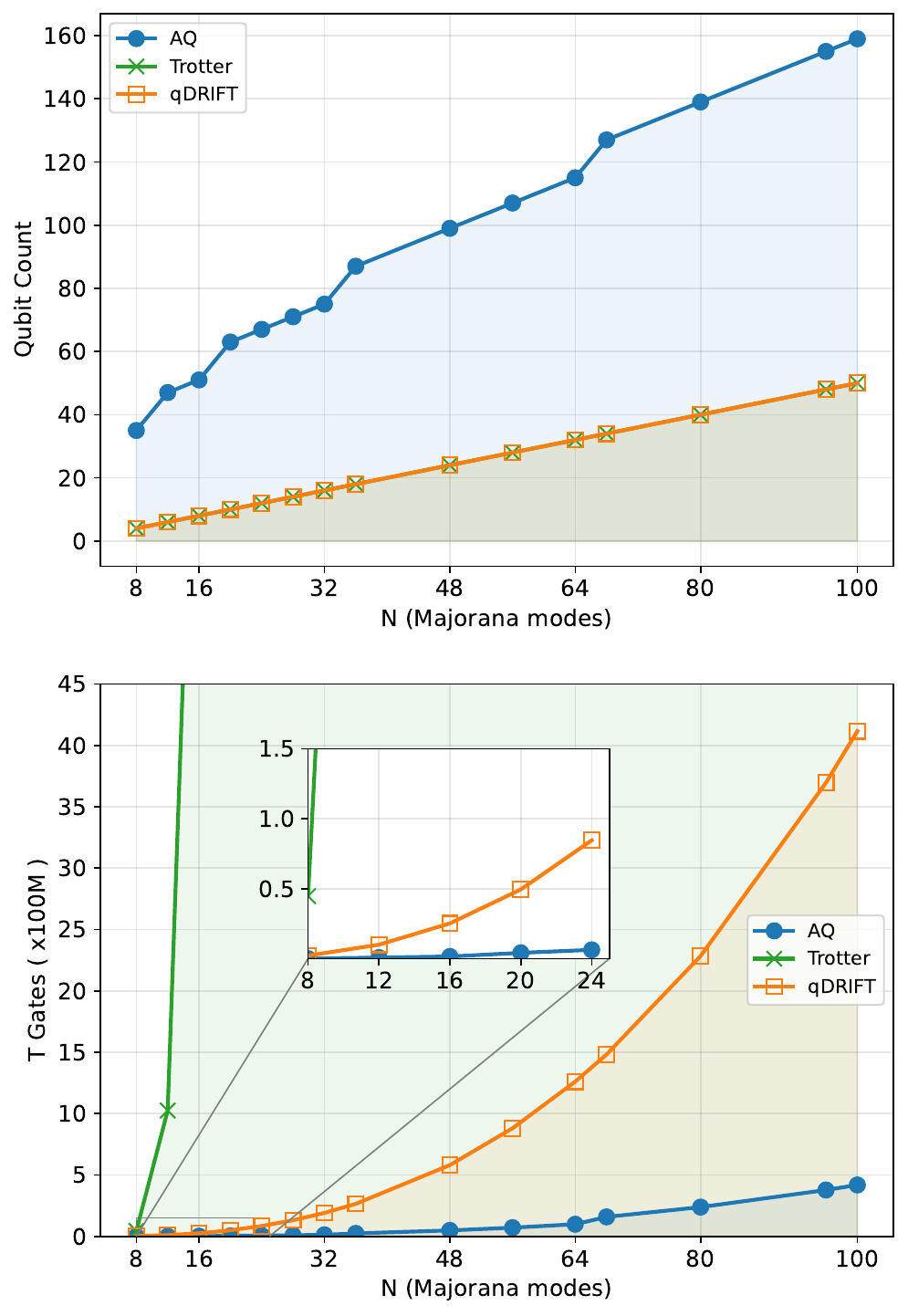}
    \caption{Qubit counts and total T gates as a function of the number of Majorana modes. The top plot shows the scaling in qubit count, while the bottom plot gives the scaling for the total number of T gates required. PPR based methods provide a better qubit count, but their T count (with qDRIFT requiring fewer resources) is much higher than the one for AQ. For the T gate scaling $\varepsilon_\mathrm{total}$ is fixed at $10^{-4}$. Coupling constants were again fixed equivalently for all methods ($Jt = 24$ for Trotter and qDRIFT and $Jt=1$ for AQ).}
    \label{fig:Nscaling}
\end{figure}

\subsection{Scaling $\varepsilon$ - Precision}
A known benefit of using qubitization and QSP for Hamiltonian simulation is the ability to scale precision for minimal cost, $\log(\frac{1}{\varepsilon})$ \cite{qubitization}. Figure \ref{fig:Escaling}, utilizing log-log scaling, highlights this property of QSP while providing a stark contrast with the PPR based methods. The second order Trotter rotations, and therefore T gates, are expected to grow at an absolute rate of $\varepsilon^{\frac{1}{\mathrm{order}}}$\cite{trotter_error} when scaling $\varepsilon$, which does match the resulting absolute slope on the log scale of $0.52$. Similarly, the qDRIFT results show a log scale absolute slope of $1.05$ matching the expected $\frac{1}{\varepsilon}$ rate. Finally, the asymmetric qubitization technique demonstrates a near-zero slope  of $\approx 0.03$ conforming to the asymptotic cost mentioned earlier. It is interesting to see the cross-over points where one technique outperforms another. While qDRIFT has fewer total T gates at a precision of less than $10^{-3}$, it is very quickly surpassed by asymmetric qubitization due to the technique's ability to improve precision almost for free. As the precision increases, Trotter does outperform qDRIFT although the number of T gates required for either technique at that precision is beyond feasibility. Again, the results empirically confirm the performance of asymmetric qubitization as epsilon scales against the qDRIFT and Trotter results.

\subsection{Scaling N Majorana Modes}
Next, we consider the behavior of all three methods as a function of the number of Majorana modes $N$. A comparison of the qubit count as well as the number of $T$ gates can be seen in Figure \ref{fig:Nscaling}. 

For the qubit count, the two methods based on PPRs require the same amount of qubits equal to half the Majorana modes in the system. This is due to the compact version of the Jordan-Wigner transform given in Eq.~\eqref{eq: JW compact} and the fact that these methods are fairly simple and do not need auxiliary qubits. In Figure \ref{fig:Nscaling} (top) we thus see a linear increase with slope $\tfrac{1}{2}$. For asymmetric qubitization, we also observe a piecewise linear increase interrupted only by finite jumps occurring at powers of two. However, the slope and offset are higher than the other two methods. For example, simulating a system of 100 Majorana modes requires 159 qubits for asymmetric qubitization, whereas qDRIFT and Trotter require only 50.  This is due to the nature of the block encoding and the fact that the Jordan-Wigner transform used preserves the dimension of the system by utilizing only $X$ operators (see Eq.~\eqref{eq: JW asym qubit}). The jumps at powers of two can be explained by the necessary auxiliary register required for the unary iteration. In sum, the top of Figure \ref{fig:Nscaling} seems to paint a picture in favor of PPR based methods. However, the qubit count of asymmetric qubitization could be further reduced. For example, implementing the compact Jordan-Wigner transformation would reduce the number of qubits needed quite easily without adding too much complexity. Furthermore, it could be beneficial to implement a Hamiltonian with increasing indices as in Eq.~\eqref{eq: H SYK trot} leading to a reduction in the number of ancilla needed for the unary iteration.

The bottom of Figure \ref{fig:Nscaling} shows the number of T gates as a function of Majorana modes for a fixed $\varepsilon_\mathrm{alg} = \varepsilon_\mathrm{rot} = 5e{-5}$. For high qubit counts, we can immediately see that the method of asymmetric qubitization provides a significantly better T scaling. For example, simulating a system of 100 Majorana modes requires less than 500 million T gates for asymmetric qubitization, whereas for qDRIFT more than 4 billion T gates are necessary. Looking at the zoomed in overlay, we can also see that for smaller systems sizes asymmetric qubitization is still better, even though the difference is less pronounced. Additionally, we can observe the beneficial scaling of qDRIFT in comparison to Trotterization.

\section{Conclusion}
Hamiltonian simulation is recognized as a strong candidate for utilizing quantum computing, and the SYK model's properties make it a potential first use case. In this work, we implement Hamiltonian simulation for the SYK model using Trotterization, qDRIFT, and asymmetric qubitization with QSP and analyze the resource estimates to highlight the differences in the techniques and the progress of the community towards efficient Hamiltonian simulation. We provide an open-source library detailing the implementation of all three methods, along with circuit diagrams and resource estimates, to help researchers better understand the simulation techniques, as well as the SYK model. Additionally, the code repository contains a range of unit tests highlighting the rigorous requirements for validating quantum algorithms and verifying quantum software. Lastly, we showcase the importance of resource estimation as a quantum computing tool to not only provide fault-tolerant estimates, but also for comparison and progress tracking.  

\section*{Acknowledgment}
This project was conducted as part of the Quantum Open Source Foundation mentorship program that connects quantum enthusiasts with mentors from academia and industry. The authors thank Mariia Mykhailova and Sean Greenaway for their mentorship, as well as the Quantum Open Source Foundation for hosting the mentorship program. Also, we thank PsiQuantum for early access to the Construct platform, including the Workbench framework.

\bibliographystyle{IEEEtran}
\bibliography{references}

\end{document}